\documentclass[namedreferences,hyperref]{spr-sola}

\usepackage[commentmarkup=margin,authormarkup=none]{changes}
\newif\ifinternalnotes
\internalnotestrue

\DeclareRobustCommand{\waleed}[1]{}
\hypersetup{colorlinks=true, citecolor=blue, linkcolor=blue, urlcolor=blue}
\usepackage{graphicx}
\usepackage{xcolor}
\usepackage{listings}
\usepackage{adjustbox} 
\usepackage[T1]{fontenc}
\usepackage{lmodern}
\makeatletter\input{t1lmr.fd}\makeatother
\DeclareFontShape{T1}{lmr}{bx}{sc}{<->ssub*cmr/bx/sc}{}

\definecolor{revblue}{RGB}{0,0,0}
\definechangesauthor[name={Revision 1},color=black]{R1}
\setaddedmarkup{\textbf{\boldmath #1}}
\setdeletedmarkup{}
\setcommentmarkup{}
\providecommand\pgfsyspdfmark[3]{}
\newif\ifrevdraft
\makeatletter
\ifbool{Changes@optiondraft}{\revdrafttrue}{\revdraftfalse}
\makeatother
\newcommand{\revnew}{\ifrevdraft\fi}
\makeatletter
\newcommand{\revfootnote}[1]{\ifrevdraft{\def\@makefnmark{\@textsuperscript{\normalfont\bfseries\@thefnmark}}\footnote{\revnew #1}}\else\footnote{#1}\fi}
\makeatother

\definecolor{codebg}{RGB}{246,243,240}
\definecolor{codekw}{RGB}{140,30,20}
\definecolor{codestr}{RGB}{20,100,70}
\definecolor{codecmt}{RGB}{130,120,115}
\newcommand{\revlststyle}{\fontsize{7.5}{9.2}\ttfamily\revnew}

\begin{document}

\begin{frontmatter}
\title{\texttt{enCHASE}: A pipeline for the calibration and analysis of CHASE/HIS observations}

\author[addressref={aff1},email={ajmalwaleed107@gmail.com}]{\inits{W.}\fnm{Waleed}~\snm{Ajmal}\orcid{0009-0006-8222-1983}}
\author[addressref={aff3}]{\inits{F.J.}\fnm{Finlay J.}~\snm{Davis}}

\author[addressref={aff2},corref,email={apietrow@aip.de}]{\inits{A.G.M.}\fnm{Alexander~G.~M.}~\snm{Pietrow}}

\author[addressref={aff3}]{\inits{M.K.}\fnm{Malcolm~K.}~\snm{Druett}\orcid{0000-0001-9845-266X}}

\author[addressref={aff4}]{\inits{C.}\fnm{Chuan}~\snm{Li}}

\author[addressref={aff4}]{\inits{Y.}\fnm{Ye}~\snm{Qiu}}

\author[addressref={aff4}]{\inits{S.}\fnm{Shihao}~\snm{Rao}}

 \author[addressref={aff1}]{\inits{S.~A.}\fnm{Syed~Amer}~\snm{Mahmood}}

\author[addressref={aff2}]{\inits{A}\fnm{Alexandros}~\snm{Stork}}

 \address[id=aff1]{University of the Punjab, Lahore, Pakistan}    \address[id=aff4]{School of Astronomy and Space Science, Nanjing University, Nanjing 210023, China}

\address[id=aff3]{Plasma Dynamics Group, School of Electrical and Electronic Engineering, University of Sheffield, Sheffield S1 3JD, UK}

\address[id=aff2]{Leibniz-Institut f\"ur Astrophysik Potsdam (AIP),
 An der Sternwarte 16, 14482 Potsdam, Germany}

\runningauthor{Ajmal et al.}
\runningtitle{\texttt{enCHASE}: calibration of CHASE H$\alpha$ observations}

\begin{abstract}
We present \texttt{enCHASE}, an open-source pipeline for the calibration and analysis of H$\alpha$ and Fe\,\textsc{i}~6569~\AA\ observations from the Chinese H$\alpha$ Solar Explorer (CHASE), addressing the lack of a public, documented tool for correcting residual instrumental effects in the mission's Level-1 data. The pipeline performs full-disk spatial registration, sub-field stabilization via correlation tracking and optical flow, {wavelength offset correction}, radiometric calibration against a high-resolution reference atlas{, and a stray-light correction.}

{We demonstrate the method on six regions of different activity within one full-disk time series, including quiet Sun, a filament, a sunspot, a plage, a prominence, and an X1.2 flare. For reductions with both lines, we achieve a sub-pixel residual pointing error for tracked sub-fields, a wavelength zero-point that is stable to 1.2~m\AA\ (0.05~km\,s$^{-1}$), and an intensity calibration uncertainty of about 1\% of the continuum. When using only H$\alpha$, we reach slightly higher uncertainties and therefore recommend the use of both lines in the reduction.}

\end{abstract}

\end{frontmatter}

\section{Introduction}
\label{sec:intro}
The Chinese H$\alpha$ Solar Explorer \citep[CHASE;][]{li2022chase},
launched in October 2021, provides full-disk spectroscopic observations
of the solar photosphere and chromosphere in the H$\alpha$ 6562.8~\AA\ and Fe\,\textsc{i} 6569.2~\AA\ lines. Since its launch, CHASE has been widely used to investigate phenomena including solar flares \citep[][]{Li2023, Li2024, Ma2024}, filament and prominence eruptions \citep[][]{Qiu2024, Liu2025}, {dynamics of the solar atmosphere} \citep[][]{Rao2024, Yu2024, Wang2026}, {and Sun-as-a-star studies} \citep[][]{Liu2025, Xie2026}.

Its H$\alpha$ Imaging Spectrograph is a slit-based spectrograph with a resolving power of approximately \(R\sim 90\; 000\){, corresponding to a Gaussian instrumental profile with an FWHM of 0.072~\AA\ \citep{li2022chase,qiu2022calibration}}. The instrument obtains a complete raster scan of the solar disk in approximately one minute, while observations of smaller regions of interest are also possible upon request. Its observing principle is broadly similar to that of ground-based slit-scanning instruments such as the Solar Explorer \citep[Sol'Ex;][]{Buil2023,Pal2025,Champeau2026}. However, its Sun-synchronous orbit, similar to those of the Interface Region Imaging Spectrograph \citep[IRIS;][]{Bart2014} and Hinode \citep{Kosugi2007}, allows for near-continuous observations of the Sun, which are currently limited to {about 30 minutes of observations per 95-minute orbit} due to downlink volume constraints.

CHASE observations are publicly available through the Solar Science Data Center (SSDC) of Nanjing University\footnote{\url{https://ssdc.nju.edu.cn/}} as Level-1 data.  {The early data on the website (from 2021 October 14 to 2023 July 18) have} undergone standard instrumental corrections such as dark and flat-field subtraction, {spectral curvature correction, wavelength calibration,} and reconstruction into a 3D cube \citep[x,y,$\lambda$,][]{qiu2022calibration}, but further processing is generally required before the data can be used for quantitative analysis. Residual instrumental effects can include spatial and spectral displacements, {wavelength offsets}, intensity variations, geometric distortions, and stray-light contamination, although data uploaded after {2023 July 18} is corrected for many of these issues. 

Although calibration procedures have been developed and applied by the CHASE instrument team, a complete, documented, and publicly accessible reduction pipeline suitable for general users is not currently available. In this work, we present the initial release of \texttt{enCHASE}, a pipeline designed to provide a reproducible workflow for reducing, calibrating, and preparing CHASE observations for scientific analysis. Each step is discussed individually in the sections below {along with the uncertainties and limitations of each step}.

\section{Example Observation}
\label{sec:example}

\begin{figure*}
    \centering
    \includegraphics[width=1\linewidth]{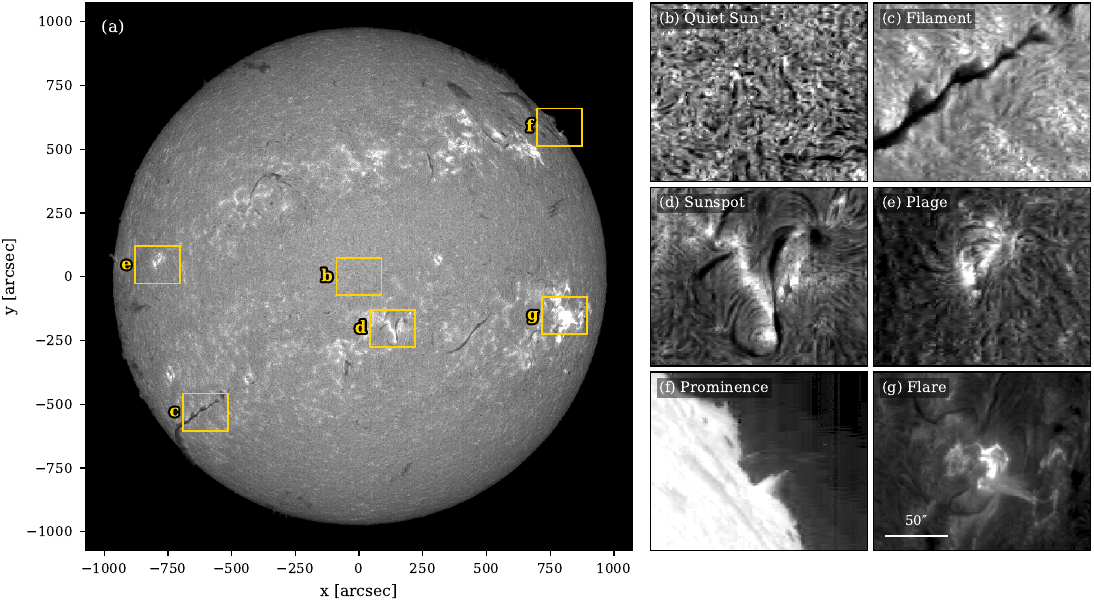}
    \caption{Overview of the Sun observed on the 29th of March, 2023, with six regions used to validate this pipeline sorted by their level of activity, from quiet Sun at disk center to a filament, a sunspot, a plage, a prominence, and a flare.}
    \label{fig:regions}
\end{figure*}

{To demonstrate the effectiveness of the enCHASE pipeline, we look at six different regions on a full-disk observation showing different levels of activity (See Fig.~\ref{fig:regions}). This time series is chosen as it contains all major issues of Level-1 CHASE data that need to be solved before one can start a scientific analysis. This includes an X1.2 flare on the X-ray classification scale of \citet{Baker1970}, which pushes the pipeline to its limits due to its rapid evolution. The analysis of these regions is outside the scope of this paper, and the data is purely used in an illustrative manner.} 

{CHASE observed the disk in its typical raster scanning mode, producing one H$\alpha$ and one Fe I cube. The native full-disk raster comprises $4625 \times 4608$ spatial pixels at 0.52 arcseconds
  per pixel, with a spectral sampling of 24.2~m\AA. The standard bin = 2 mode used here yields cubes of $2313 \times 2304$ spatial pixels at 1.04 arcseconds per pixel and a spectral sampling of 48.4~m\AA.
  The H$\alpha$ and Fe I cubes cover 6559.42--6565.09 \AA\ and 6567.52--6569.70 \AA, sampled at 118 and 46 wavelength points, respectively.}

 {The time series covers 25 scans between 02:12~UT and 02:42~UT on 2023 March 29 at a cadence of about 71~s.}

\section{Methods}
\label{sec:calibration}

In this section, we demonstrate the \texttt{enCHASE} workflow on the
example observations. We begin with data acquisition, proceed to feature
tracking and stabilization, and conclude with the {wavelength offset correction, the intensity calibration, and the stray-light correction}.

\subsection{Data Acquisition}
\label{sec:download}
Level-1 CHASE data can be obtained in three ways. The most common route is to download files interactively from the SSDC portal, which issues signed, time-limited download links one file at a time. These links can also be passed to \texttt{enCHASE} directly, either as a single URL or as a text file of links, which are then downloaded in a resumable way. If the data already exists locally, the code can simply be pointed to the folder. Finally, the portal client packaged with the pipeline offers a fully programmatic route. After a one-time portal login, it searches the archive and generates download links for any time range from the command line.

\begin{lstlisting}[language=bash,caption={Programmatic data access. Signed
links expire after $\sim$24\,h, so they are minted immediately before
downloading.},label={lst:portal}]
chase-portal login   # once
chase-portal urls --start "2023-03-29 02:00" \
    --end "2023-03-29 03:00" -o links.txt
chase links.txt      # download + calibrate
\end{lstlisting}

\subsection{Spatial Calibration}
\label{sec:spatial}

\begin{figure}

\centerline{\includegraphics[width=\textwidth,clip=]{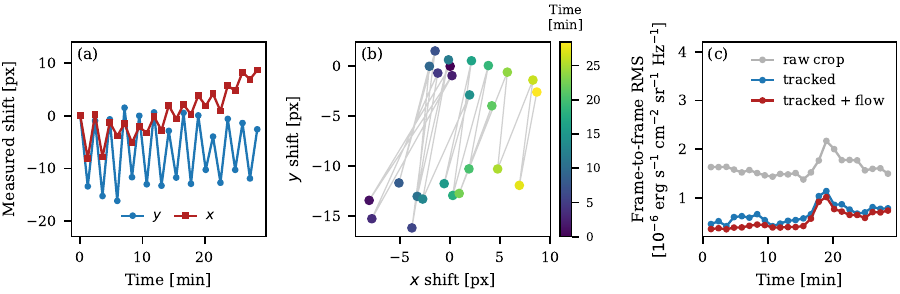}}
\caption[Spatial calibration]{Spatial calibration of the example observation. (a)~Measured $x$ (red) and $y$ (blue) shifts per scan. The $y$ shifts alternate by 6--18\,px with the raster direction, while $x$ drifts from $-8$ to $+9$\,px over the sequence. (b)~The same shifts in the $x$--$y$ plane, colored by time. (c)~{RMS intensity difference between consecutive scans at 6565.1~\AA} before correction (gray), after tracking (blue), and after adding optical flow (red){, converted to physical units with the intensity factor of Fig.~\ref{fig:atlas}}. The peak near $t = 18$\,min coincides with the flare rise.}
\label{fig:chain}
\end{figure}

\begin{figure}
\centerline{\includegraphics[width=\textwidth,clip=]{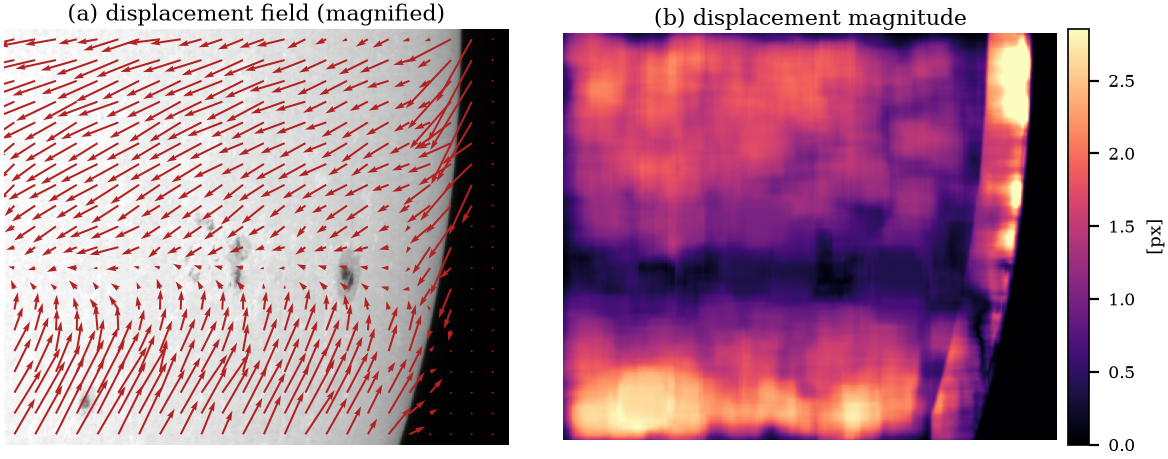}}
\caption{The optical-flow field between two consecutive scans, computed on {a reference wavelength that is only weakly affected by the flare}. (a) The displacement vectors, magnified for
visibility, overplotted on the first frame. (b) The magnitude of the same
field shown as an intensity map.}
\label{fig:flowfield}
\end{figure}

Spatial calibration operates at two levels. At the full-disk level, the disk center of each scan is located by a Hough-transform circle fit to the limb \citep{Kimme1975}, allowing frames to be recentered to the nearest pixel, and the
\texttt{INST\_ROT} header angle allows each frame to be derotated to solar north. Together, these functions register full-disk images across each scan. {Alternatively, a least-squares fit of a circle or rotated ellipse to the limb provides sub-pixel recentering of the frames.}

The magnitude of these movements is shown in Fig.~\ref{fig:chain} where the first two panels show the raw offsets found for each frame, and the third panel shows {the root-mean-square (RMS) intensity difference between consecutive scans, computed over all pixels of the H$\alpha$ far-wing image at 6565.1~\AA, which drops after each correction step.}

\subsubsection{Correlation tracking}
When looking at smaller regions of interest (ROIs), these can be tracked and stabilized across scans using  phase cross-correlation
\citep{guizar2008efficient}. This method works best when dealing with slowly changing frames, and therefore the far wings are selected to avoid rapidly evolving chromospheric structures. This method does not distort the maps, and only removes global movement patterns.

Since CHASE scans in a back-and-forth alternating fashion, consecutive scans tend to have a pointing offset, while scans taken in the same direction align better (see Fig.~\ref{fig:chain}, panels a and b). This correlation is leveraged by applying the code to the series as a whole, fitting a Theil--Sen trend \citep{theil1950rank} to the scans of each direction separately, which allows for the interpolation of outlier positions. {A shift that deviates from the trend of its raster direction by more than 4~px (4.2~arcsec on the $2\times2$-binned grid) is treated as a failed correlation and replaced by interpolation between the other scans of that direction. The reference scan is excluded from this test, given that its shift is zero by construction. Thresholds from 0.5 to 15~px were tested on the flare field: 3--15~px replace no shifts, while lower values also replace valid ones.}

\begin{lstlisting}[basicstyle=\revlststyle,caption={Robust tracking. \texttt{robust\_shifts} rejects
failed correlations without disturbing genuine drift.{\revnew\ Shifts more than \texttt{threshold} pixels (default 4.0) from the Theil--Sen trend of their raster direction are replaced.}},label={lst:track}]
seq = chase.load_flare_sequence(
    "./data", patch=[940, 1080, 1870, 2040],
    smooth_shifts=True)   # default: robust_shifts(threshold=4.0)
# X1.2: no scan deviates from its parity trend by more than 4 px
\end{lstlisting}

\subsubsection{Optical-Flow Stabilization}
Following the corrections of the global movements and solar rotation, we still find localized movements in the data which are likely caused by vibrations in the telescope while it is scanning. These vibrations introduce a seeing-like distortion to the data which needs to be corrected on a frame-by-frame basis.

These residual rubber-sheet distortions, see Fig.~\ref{fig:flowfield}, are removed using dense optical flow, computed with the polynomial-expansion method of \citet{farneback2003two}. The flow is measured on {a reference wavelength that is only weakly affected by the flare (the Fe\,\textsc{i} background at 6569.7~\AA)} and applied identically to every wavelength of both cubes, under the assumption that these distortions are instrumental and therefore independent of
wavelength. {The red curve in Fig.~\ref{fig:chain}c shows the frame-to-frame RMS after optical flow.}

\subsection{Wavelength Offset Correction}
\label{sec:wavelength}

The wavelength zero-point of the CHASE Level-1 files drifts from scan to scan. Given the steep slope of the H$\alpha$ wings, this drift alone can produce intensity differences comparable to those of a small flare when spectra from different scans are compared channel by channel. In \texttt{enCHASE}, the wavelength axis of each file is therefore reconstructed from its own \texttt{CRVAL3} and \texttt{CDELT3} header keywords. Every cube is then resampled onto the wavelength grid of the first scan by cubic interpolation.

\begin{lstlisting}[basicstyle=\revlststyle,caption={Wavelength alignment. Both corrections are switchable for users who require untouched spectra.},label={lst:wave}]
fixed = chase.resample_wavelength(cube, src_wav, dst_wav, kind="cubic")
corrected, shifts = chase.correct_spectral_drift(cubes)
# shifts are in channels (sub-pixel)
\end{lstlisting}

\begin{figure}
\revnew
\centerline{\includegraphics[width=\textwidth,clip=]{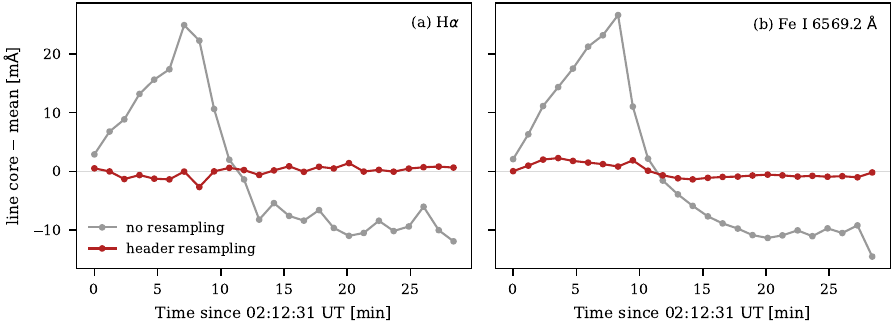}}
\caption[Line-core position per scan]{{\revnew Line-core position of (a) H$\alpha$ and (b) Fe\,\textsc{i} 6569.2~\AA\ in the $100\times100$~px quiet-Sun box at disk center, relative to the mean over the 25 scans, without resampling (gray) and after resampling onto the header wavelength grid (red).}}
\label{fig:val_wave}
\end{figure}

{Fig.~\ref{fig:val_wave} shows the scan-to-scan scatter of the line-core position in the disk-center box before and after resampling. The resampling reduces this scatter from 11.6~m\AA\ (H$\alpha$) and 12.7~m\AA\ (Fe\,\textsc{i}) to 0.9 and 1.2~m\AA\ (0.04 and 0.05~km\,s$^{-1}$), which is comparable to or below the uncertainty of a single centroid measurement (2.0~m\AA\ for H$\alpha$ and 1.1~m\AA\ for Fe\,\textsc{i}).}

\subsection{Intensity Calibration}
\label{sec:intensity}

\begin{figure}
\centerline{\includegraphics[width=0.8\textwidth,clip=]{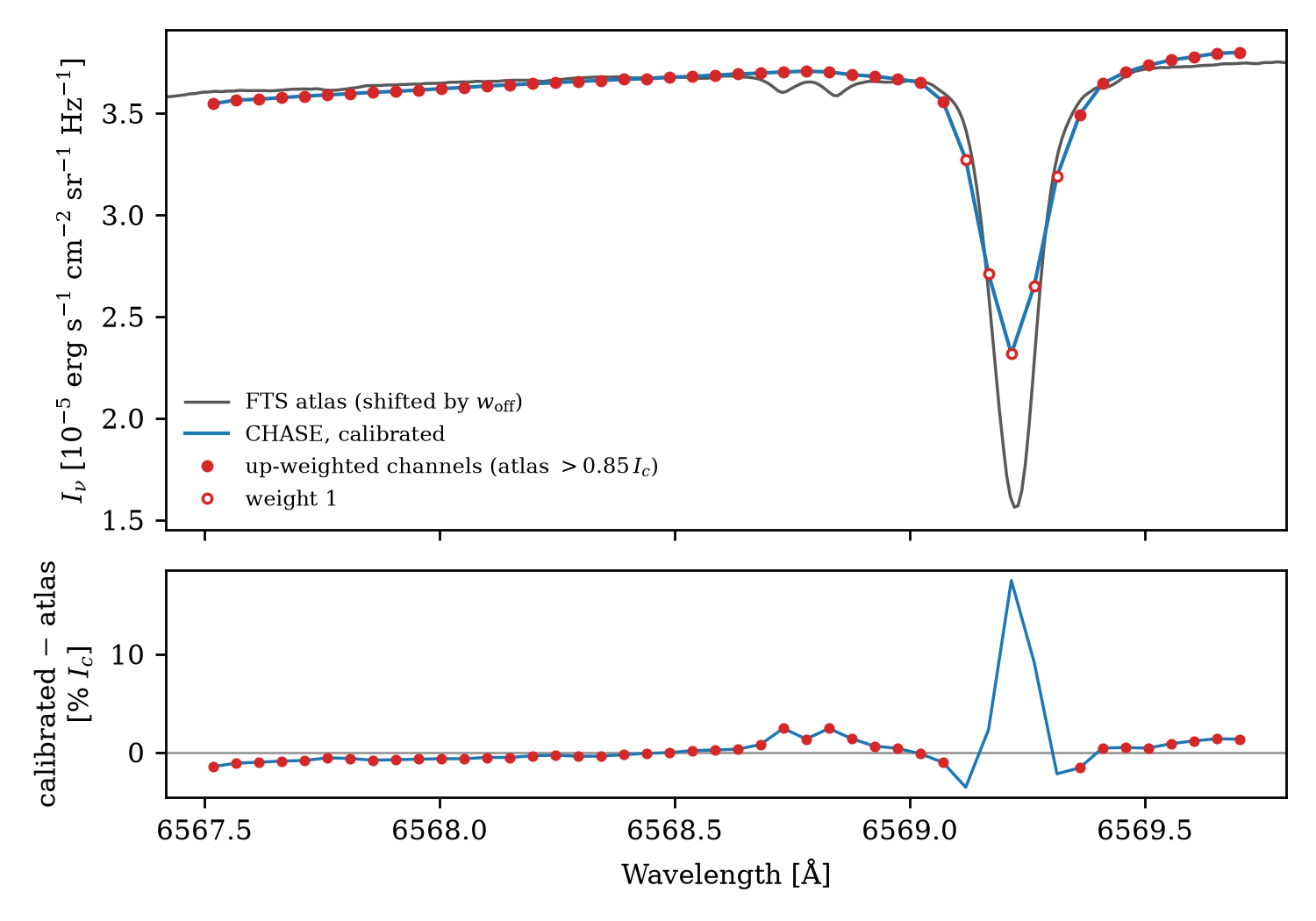}}
\caption[Atlas calibration]{{Atlas calibration of the last scan (Listing~\ref{lst:atlas}). The top panel shows the calibrated mean Fe\,\textsc{i} profile of the disk-center box (blue) and the FTS atlas shifted by the fitted wavelength offset (gray), where filled points mark the wing channels, which receive a 20 times higher weight in the fit. The bottom panel shows the difference between the two in percent of the continuum. The fitted intensity factor is $2.303\times10^{-8}$~erg\,s$^{-1}$\,cm$^{-2}$\,sr$^{-1}$\,Hz$^{-1}$\,DN$^{-1}$. The large residual in the line core arises because the atlas is not convolved with the instrumental profile and the stray light is not removed at this stage (Sect.~\ref{sec:stray}).}}
\label{fig:atlas}
\end{figure}

{Intensity calibration is performed on a $100\times100$~px box at disk center in the Fe\,\textsc{i} cube, whose mean spectrum is fitted to the solar atlas of \citet{neckel1984fts} with ISPy \citep{ispy2021},} yielding an intensity factor and a wavelength offset. {The calibration is carried out separately for each data set, meaning that any loss of throughput since launch is absorbed into the calibration factor of that data set rather than propagating into the calibrated intensities. The box is centered on the disk center by default, and can be placed by hand to avoid active regions. Optionally, remaining active pixels can be excluded, keeping only pixels within 5\% of the box median in the continuum and 10\% in the line core (Listing~\ref{lst:atlas}).  {Since at this stage the instrumental profile has not been deconvolved and the stray light has not yet been corrected,} the line core cannot be matched to the unconvolved atlas, and the fit is therefore weighted towards the wings. The 85\% threshold selects the atlas channels that receive a 20 times higher weight in the fit (Fig.~\ref{fig:atlas}), and is applied to the wavelength channels and not to the image pixels.}

\begin{lstlisting}[basicstyle=\revlststyle,caption={Absolute intensity calibration against the \citet{neckel1984fts} atlas.},label={lst:atlas}]
dc_cube, dc_wav, _ = chase.extract_disk_center("RSM20230329T024055_0024_FE.fits")
ifact, woff = chase.atlas_calibrate(dc_wav, dc_cube,
    wing_continuum_frac=0.85,  # atlas channels, not pixels
    mask_active=False)         # True: drop active pixels
\end{lstlisting}

{For the data set, the fit reproduces the Fe\,\textsc{i} wings to 0.95\% RMS of the continuum, with a largest residual of 2.4\% (Fig.~\ref{fig:atlas}), and the factor varies by 0.09\% between scans. Thresholds of 0.70--0.85 change the factor by at most 0.15\% (2.4~K in brightness temperature), 0.90 by 0.8\%, and the pixel mask by less than 0.05\%. After limb-darkening correction and without the mask, moving the box by up to 100~px, resizing it to 50--200~px, or placing it on the darkest feature or brightest plage on the disk changes it by at most 0.4\%. For H$\alpha$-only data, a threshold of 0.75 gives a factor within 0.7\% of the Fe\,\textsc{i} fit.}

\subsection{Instrumental profile and stray light}
\label{sec:stray}

{The CHASE/HIS spectra contain stray light, which fills in the line cores \citep{Hong2022}. \citet{Hou2020} estimate the instrumental profile and the stray light together by comparing the disk-center profile with a reference atlas that is convolved with a Gaussian of a given FWHM, to which a constant stray-light level $\alpha$ is added. This method was applied to CHASE by \citet{Hong2022}. However, this approach requires a separate solution for each window, with a FWHM of 0.105~\AA\ and $\alpha = 10.5\%$ in Fe\,\textsc{i}, but 0.140~\AA\ and $\alpha = 5.0\%$ in H$\alpha$, even though both windows are recorded simultaneously on the same detector and should therefore share one nearly identical instrumental profile. We instead choose to describe both windows with a single profile, in which the constant veil is replaced by a second, broad Gaussian that carries a fraction $f$ of the light,}
\begin{equation}
  \chi^2(w, W, f) \;=\; \Big[\, I_\lambda^{\mathrm{obs}} - \Big( (1-f)\, I_\lambda^{\mathrm{FTS}} \otimes G(w) \;+\; f\, I_\lambda^{\mathrm{FTS}} \otimes G(W) \Big) \Big]^2 ,
  \label{eq:stray}
\end{equation}
{where $\otimes$ represents the convolution, $I_\lambda^{\mathrm{obs}}$ is the mean profile of a $51\times51$~px box at disk center in the first scan, normalized to the continuum with a low-order polynomial that is fitted together with a wavelength offset, $I_\lambda^{\mathrm{FTS}}$ the atlas of \citet{neckel1984fts} through \citet{Pietrow2026hs}, and $G(w)$ and $G(W)$ are Gaussians with a FWHM of $w$ and $W$. With $w = 0.100$~\AA, $W = 1.8$~\AA\ and $f = 14.5\%$, this single set of parameters reproduces the disk-center profile to 0.3\% RMS of the continuum in Fe\,\textsc{i} and 0.6\% in H$\alpha$, compared with 0.3\% and 0.5\% for the two separate solutions used by \citet{Hong2022}. The fitted core is broader than the nominal Gaussian instrumental profile with a FWHM of 0.072~\AA\ \citep{li2022chase,qiu2022calibration}.}

\begin{figure}
\revnew
\centerline{\includegraphics[width=\textwidth,clip=]{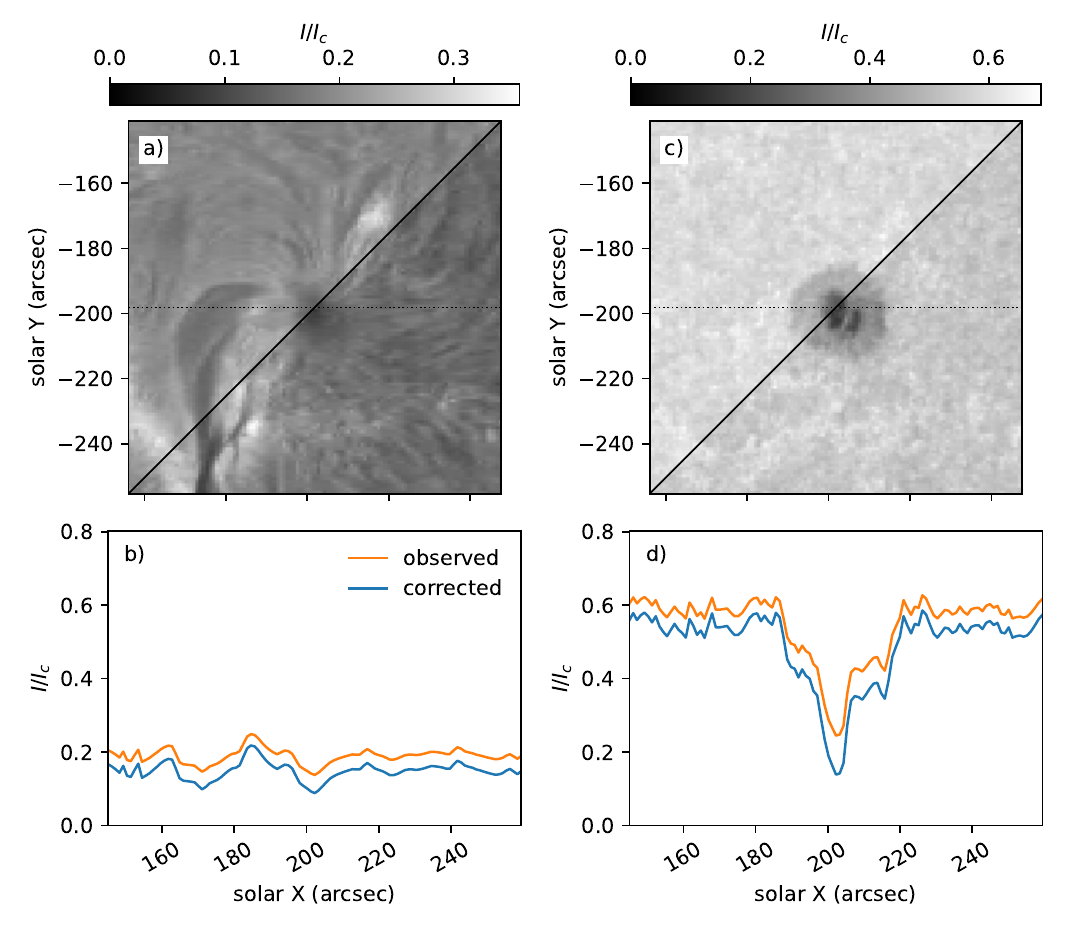}}
\caption[Stray-light correction]{{\revnew Effect of the stray-light correction on the sunspot of Fig.~\ref{fig:regions}. (a) and (c) show the spot in the H$\alpha$ and Fe\,\textsc{i} 6569 line cores, cut along the diagonal, as observed above the line and corrected below it. (b) and (d) give the intensity along the dotted row, observed in orange and corrected in blue.}}
\label{fig:straylight}
\end{figure}

{Following \citet{Hong2022}, the stray light is assumed to be the same over the full field of view. The broad term is therefore taken from the disk-center profile $I_\lambda^{\mathrm{DC}}$, and every profile is corrected as}
\begin{equation}
  I_\lambda \;=\; \frac{I_\lambda^{\mathrm{obs}} - f\, I_\lambda^{\mathrm{DC}} \otimes G(W)}{1-f}.
  \label{eq:straycorr}
\end{equation}
{This removes the broad wings of the instrumental profile, while its narrow core is not deconvolved. Fig.~\ref{fig:straylight} shows the effect on the sunspot of Fig.~\ref{fig:regions}. In the quiet surroundings the continuum is left unchanged to within 1\%, while the line cores deepen by 22\% in H$\alpha$ and 8\% in Fe\,\textsc{i}. The umbra darkens at all wavelengths, so that its core intensity relative to these surroundings drops from 0.41 to 0.25 in Fe\,\textsc{i} and from 0.74 to 0.61 in H$\alpha$, and its continuum contrast from 0.37 to 0.27.}

{A stray-light level that is constant over the field of view is an approximation. It is set at disk center and therefore overestimates the stray light close to and beyond the limb, where the disk is fainter, meaning that this correction should not be applied to off-limb features such as prominences. In addition, light scattered from the immediate surroundings of a pixel is not accounted for, given that no spatial point spread function of CHASE is currently available.}

\section{Results and discussion}
\label{sec:products}

After applying the steps described above, we are left with a processed time series, either of a sub-field or of the full disk, with most instrumental effects removed. A clean way of showing the result of these post-processing steps is by means of contrast profiles \citep[e.g.,][]{DeWilde2025}, where the change in spectral shape is captured as a function of time.

{For quiet patches we expect little to no change in such maps, while rapidly evolving regions such as flares show specific signals. We define the contrast as}
\begin{equation}
  C(\lambda, t) \;=\;
   \frac{I_\mathrm{ROI}(\lambda, t) - I_\mathrm{QS}(\lambda, t)}{I_\mathrm{QS}(\lambda, t)}   \;-\;\Big\langle \frac{I_\mathrm{ROI} - I_\mathrm{QS}}{I_\mathrm{QS}} \Big\rangle_{t},
  \label{eq:contrast}
\end{equation}
where $I_\mathrm{QS}$ is {the $100\times100$~px quiet-Sun box at disk center} and $\langle\cdot\rangle_t$ is the average over all 25 scans.

\begin{figure}
\centerline{\includegraphics[width=0.8\textwidth,clip=]{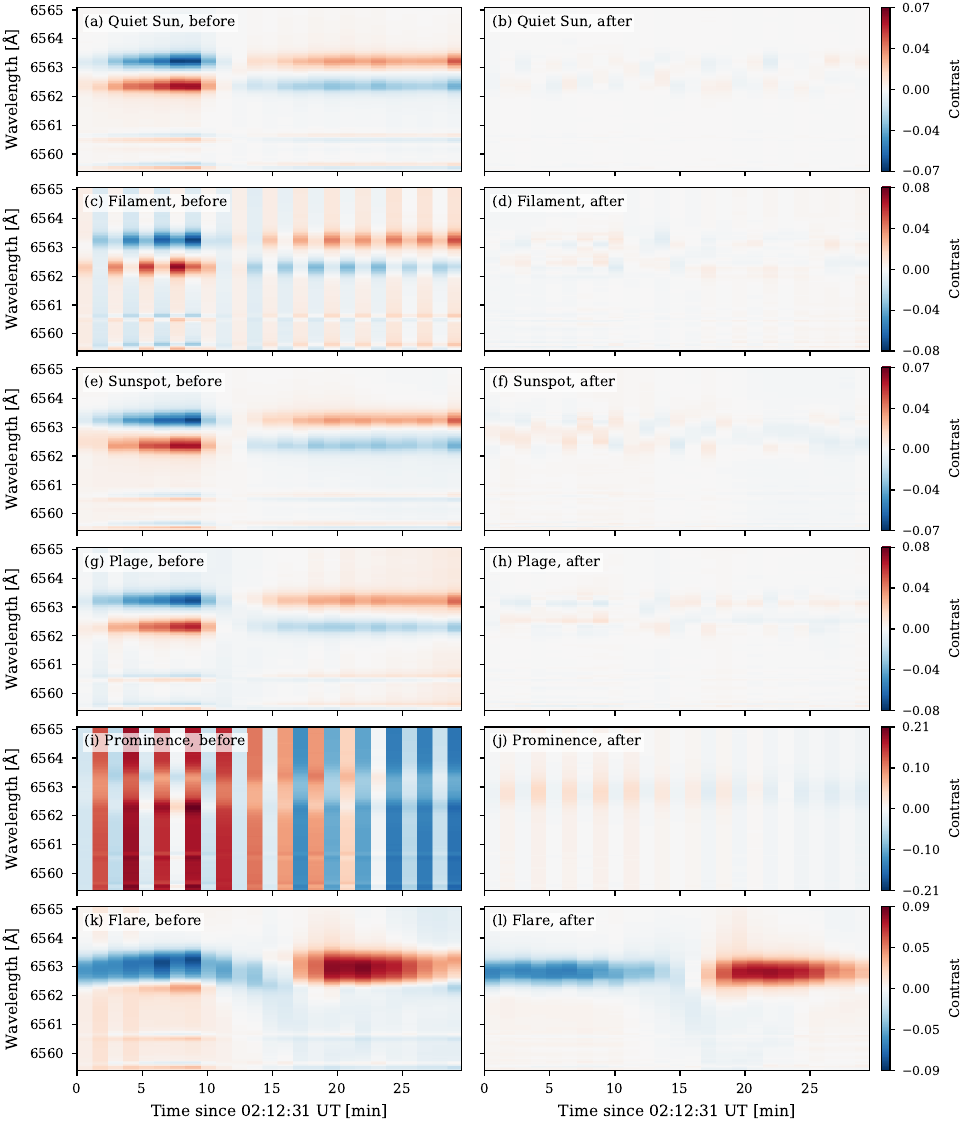}}
\caption{{Contrast (Equation~\ref{eq:contrast}) as a function of wavelength and time for the six regions of Fig.~\ref{fig:regions}. Left: fixed crop of the Level-1 data with no tracking and no wavelength correction, relative to the first scan. Right: \texttt{enCHASE}-calibrated data, relative to a quiet-Sun box at disk center. In both columns the time average of each wavelength is subtracted. Each row has its own linear color scale, and no value is clipped. The flare row shows the line-core brightening at scan 18 (02:33~UT), with a peak contrast of 0.08. The stripes left in the prominence row alternate with the raster direction.}}
\label{fig:contrast}
\end{figure}

{The contrast before and after calibration is shown in Fig.~\ref{fig:contrast} for each of the six regions of Fig.~\ref{fig:regions}. Before calibration, every region shows similar patterns, which are caused by the drift of the wavelength zero-point and the pointing. After calibration, the quiet Sun, the filament, the sunspot, and the plage are flat to about 1\%, with only the flare remaining with a line-core brightening that peaks at scan 18 (02:33~UT) with a contrast of 0.08. In the prominence, stripes remain that alternate with the raster direction, because correlation tracking cannot lock onto the off-limb field.}

\begin{figure}
\revnew
\centerline{\includegraphics[width=\textwidth,clip=]{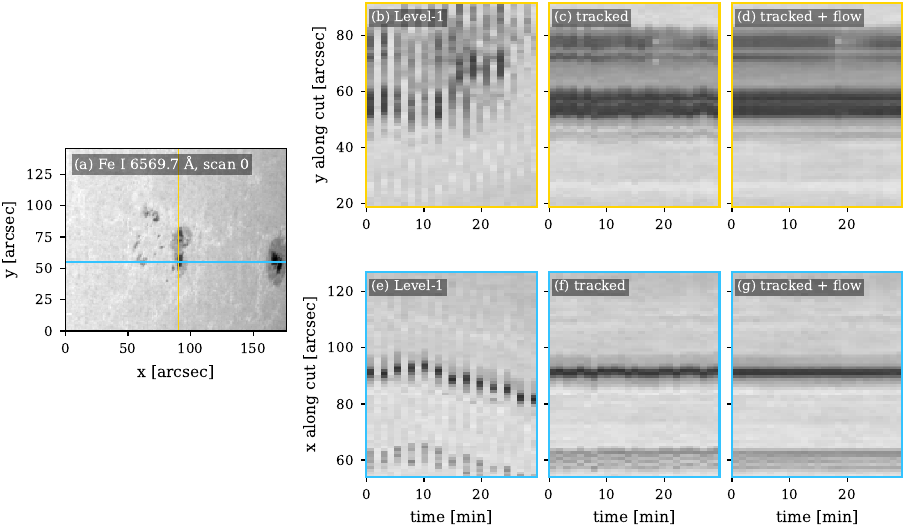}}
\caption{{\revnew Time slices through the sunspot in the flare region, in Fe\,\textsc{i} 6569.7~\AA. (a) First scan with the two cuts through the umbra. (b)\,--\,(d) Intensity along the vertical cut as a function of time, for the fixed crop of the Level-1 data, after correlation tracking, and after tracking and optical flow. (e)\,--\,(g) The same for the horizontal cut.}}
\label{fig:timeslice}
\end{figure}

{This is further illustrated in Fig.~\ref{fig:timeslice}, where a time series of two simulated slits is shown before and after correction. In the Level-1 data the umbra jumps between consecutive scans and drifts over the sequence. After tracking and optical flow it stays in place. Here we find that the residual pointing error drops from 10.5~arcsec RMS to 0.43~arcsec after tracking and to 0.06~arcsec after optical flow, well below one pixel.}

\section{Conclusions}
\label{sec:discussion}
We have presented \texttt{enCHASE}, an open-source pipeline for the post-processing of CHASE/HIS Level-1 data into science-ready spectral cubes through full-disk registration, sub-field stabilization, wavelength alignment, and absolute intensity calibration against the atlas of
\citet{neckel1984fts}{, and a stray-light correction}.

As an illustrative example, we applied the pipeline to a {29-minute time series of 25 scans} covering a wide range of solar features including an eruptive X1.2 flare. The Level-1 data contained large jitter patterns in pointing and {wavelength}, both of which our pipeline can correct (See Fig.~\ref{fig:chain}), as well as the smaller-scale rubber-sheet distortions (see Fig. \ref{fig:flowfield}). {Additionally, {an intensity calibration and wavelength offset correction were applied} to the data. Together with the tracking and optical flow, this greatly reduced the frame-to-frame RMS of the time series (see Fig.~\ref{fig:chain}c), leaving mainly the solar signal rather than instrumental jitter.} {After calibration, the pointing error of a tracked sub-field is well below one pixel (0.06~arcsec) and the wavelength zero-point is stable to about 1~m\AA\ (0.05~km\,s$^{-1}$), and the quiet Sun, filament, sunspot, and plage of Fig.~\ref{fig:regions} are flat to about 1\% in contrast.} {The stray light is estimated at disk center with a modified version of the method of \citet{Hou2020}, in which a single instrumental profile with a broad component that carries 14.5\% of the light describes both windows, and is removed assuming that it is constant across the field of view \citep{Hong2022}. Currently no spatial point spread function of CHASE has been published, so light scattered from the immediate surroundings of a pixel is not corrected, and the correction should not be applied to off-limb features.} The \texttt{enCHASE} library  is distributed on PyPI as {\texttt{enchase}}\revfootnote{\url{https://pypi.org/project/enchase/}} (version {2.2.0} at the time of writing) under the MIT license. {The source code is shared on the project GitHub}\revfootnote{\url{https://github.com/waleedsworld/enchase}}.

\begin{acks}

AP is supported by the \emph{Deut\-sche For\-schungs\-ge\-mein\-schaft, DFG\/} project number PI 2102/1-1.
MD was supported by UK
Research and Innovation (UKRI) [grant number UKRI1165].
{The CHASE mission is supported by China National Space Administration. CHASE team of NJU is supported by NSFC (12333009) and the Fundamental Research Funds for the Central Universities (KG202506).}

\end{acks}

\begin{dataavailability}
The Level-1 CHASE observations used in this paper are publicly available from the Solar Science Data Center of Nanjing
University at \url{https://ssdc.nju.edu.cn/}. 
\end{dataavailability}

\begin{codeavailability}
\texttt{enCHASE} is open source under the MIT license
({\url{https://github.com/waleedsworld/enchase}}) and installable from PyPI
({\texttt{pip install enchase}, version 2.2.0}).
\end{codeavailability}

\begin{ethics}
\begin{conflict}
The authors declare that they have no conflicts of interest.
\end{conflict}
\end{ethics}

\bibliographystyle{spr-mp-sola}
\bibliography{references}

@ARTICLE{li2022chase,
  author = {{Li}, Chuan and {Fang}, Cheng and {Li}, Zhen and {Ding}, Mingde and
            {Chen}, Pengfei and {Qiu}, Ye and {You}, Wei and {Yuan}, Yuan and
            {An}, Minjie and {Yu}, Hongjiang and {Li}, Wei},
  title = {The {Chinese} {H}$\alpha$ {Solar} {Explorer} ({CHASE}) mission: An overview},
  journal = {Science China Physics, Mechanics, and Astronomy},
  year = 2022,
  volume = {65},
  number = {8},
  pages = {289602},
  doi = {10.1007/s11433-022-1893-3},
}

@ARTICLE{Hou2020,
       author = {{Hou}, Jun-Feng and {Xu}, Zhi and {Yuan}, Shu and {Chen}, Yu-Chao and {Peng}, Jian-Guo and {Wang}, Dong-Guang and {Xu}, Jun and {Deng}, Yuan-Yong and {Jin}, Zhen-Yu and {Ji}, Kai-Fan and {Liu}, Zhong},
        title = "{Spectro-polarimetric observations at the NVST: I. instrumental polarization calibration and primary measurements}",
      journal = {Research in Astronomy and Astrophysics},
         year = 2020,
        month = apr,
       volume = {20},
       number = {4},
          eid = {045},
        pages = {045},
          doi = {10.1088/1674-4527/20/4/45},
archivePrefix = {arXiv},
       eprint = {1911.07397},
 primaryClass = {astro-ph.SR},
       adsurl = {https://ui.adsabs.harvard.edu/abs/2020RAA....20...45H}
}

@ARTICLE{Pietrow2026hs,
       author = {{Pietrow}, Alexander G.~M.},
        title = "{HelioSpectrotron 5000: an interactive solar atlas}",
      journal = {The Open Journal of Astrophysics},
         year = 2026,
        month = feb,
       volume = {9},
        pages = {58273},
          doi = {10.33232/001c.158273},
archivePrefix = {arXiv},
       eprint = {2602.20101},
 primaryClass = {astro-ph.SR},
       adsurl = {https://ui.adsabs.harvard.edu/abs/2026OJAp....958273P}
}

@ARTICLE{DeWilde2025,
       author = {{De Wilde}, M. and {Pietrow}, A.~G.~M. and {Druett}, M.~K. and {Pastor Yabar}, A. and {Koza}, J. and {Kontogiannis}, I. and {Andriienko}, O. and {Berlicki}, A. and {Brunvoll}, A.~R. and {de la Cruz Rodr{\'\i}guez}, J. and {Thoen Faber}, J. and {Joshi}, R. and {Kuridze}, D. and {N{\'o}brega-Siverio}, D. and {Rouppe van der Voort}, L.~H.~M. and {Ryb{\'a}k}, J. and {Scullion}, E. and {Silva}, A.~M. and {Vashalomidze}, Z. and {Vicente Ar{\'e}valo}, A. and {Wi{\'s}niewska}, A. and {Yadav}, R. and {Zaqarashvili}, T.~V. and {Zbinden}, J. and {{\O}yre}, E.~S.},
        title = "{Synthesizing Sun-as-a-star flare spectra from high-resolution solar observations}",
      journal = {A\&A},
         year = 2025,
        month = aug,
       volume = {700},
          eid = {A275},
        pages = {A275},
          doi = {10.1051/0004-6361/202554870},
archivePrefix = {arXiv},
       eprint = {2507.07967},
 primaryClass = {astro-ph.SR},
       adsurl = {https://ui.adsabs.harvard.edu/abs/2025A&A...700A.275D}
}

@article{Kimme1975,
  title = {Finding circles by an array of accumulators},
  volume = {18},
  ISSN = {1557-7317},
  url = {http://dx.doi.org/10.1145/360666.360677},
  DOI = {10.1145/360666.360677},
  number = {2},
  journal = {Communications of the ACM},
  publisher = {Association for Computing Machinery (ACM)},
  author = {Kimme,  Carolyn and Ballard,  Dana and Sklansky,  Jack},
  year = {1975},
  month = Feb,
  pages = {120–122}
}

@ARTICLE{Kosugi2007,
       author = {{Kosugi}, T. and {Matsuzaki}, K. and {Sakao}, T. and {Shimizu}, T. and {Sone}, Y. and {Tachikawa}, S. and {Hashimoto}, T. and {Minesugi}, K. and {Ohnishi}, A. and {Yamada}, T. and {Tsuneta}, S. and {Hara}, H. and {Ichimoto}, K. and {Suematsu}, Y. and {Shimojo}, M. and {Watanabe}, T. and {Shimada}, S. and {Davis}, J.~M. and {Hill}, L.~D. and {Owens}, J.~K. and {Title}, A.~M. and {Culhane}, J.~L. and {Harra}, L.~K. and {Doschek}, G.~A. and {Golub}, L.},
        title = "{The Hinode (Solar-B) Mission: An Overview}",
      journal = {\solphys},
         year = 2007,
        month = jun,
       volume = {243},
       number = {1},
        pages = {3-17},
          doi = {10.1007/s11207-007-9014-6},
       adsurl = {https://ui.adsabs.harvard.edu/abs/2007SoPh..243....3K}
}

@ARTICLE{Bart2014,
       author = {{De Pontieu}, B. and {Title}, A.~M. and {Lemen}, J.~R. and {Kushner}, G.~D. and {Akin}, D.~J. and {Allard}, B. and {Berger}, T. and {Boerner}, P. and {Cheung}, M. and {Chou}, C. and {Drake}, J.~F. and {Duncan}, D.~W. and {Freeland}, S. and {Heyman}, G.~F. and {Hoffman}, C. and {Hurlburt}, N.~E. and {Lindgren}, R.~W. and {Mathur}, D. and {Rehse}, R. and {Sabolish}, D. and {Seguin}, R. and {Schrijver}, C.~J. and {Tarbell}, T.~D. and {W{\"u}lser}, J.-P. and {Wolfson}, C.~J. and {Yanari}, C. and {Mudge}, J. and {Nguyen-Phuc}, N. and {Timmons}, R. and {van Bezooijen}, R. and {Weingrod}, I. and {Brookner}, R. and {Butcher}, G. and {Dougherty}, B. and {Eder}, J. and {Knagenhjelm}, V. and {Larsen}, S. and {Mansir}, D. and {Phan}, L. and {Boyle}, P. and {Cheimets}, P.~N. and {DeLuca}, E.~E. and {Golub}, L. and {Gates}, R. and {Hertz}, E. and {McKillop}, S. and {Park}, S. and {Perry}, T. and {Podgorski}, W.~A. and {Reeves}, K. and {Saar}, S. and {Testa}, P. and {Tian}, H. and {Weber}, M. and {Dunn}, C. and {Eccles}, S. and {Jaeggli}, S.~A. and {Kankelborg}, C.~C. and {Mashburn}, K. and {Pust}, N. and {Springer}, L. and {Carvalho}, R. and {Kleint}, L. and {Marmie}, J. and {Mazmanian}, E. and {Pereira}, T.~M.~D. and {Sawyer}, S. and {Strong}, J. and {Worden}, S.~P. and {Carlsson}, M. and {Hansteen}, V.~H. and {Leenaarts}, J. and {Wiesmann}, M. and {Aloise}, J. and {Chu}, K.-C. and {Bush}, R.~I. and {Scherrer}, P.~H. and {Brekke}, P. and {Martinez-Sykora}, J. and {Lites}, B.~W. and {McIntosh}, S.~W. and {Uitenbroek}, H. and {Okamoto}, T.~J. and {Gummin}, M.~A. and {Auker}, G. and {Jerram}, P. and {Pool}, P. and {Waltham}, N.},
        title = "{The Interface Region Imaging Spectrograph (IRIS)}",
      journal = {\solphys},
         year = 2014,
        month = jul,
       volume = {289},
       number = {7},
        pages = {2733-2779},
          doi = {10.1007/s11207-014-0485-y},
archivePrefix = {arXiv},
       eprint = {1401.2491},
 primaryClass = {astro-ph.SR},
       adsurl = {https://ui.adsabs.harvard.edu/abs/2014SoPh..289.2733D}
}

@ARTICLE{guizar2008efficient,
  author = {{Guizar-Sicairos}, Manuel and {Thurman}, Samuel T. and {Fienup}, James R.},
  title = {Efficient subpixel image registration algorithms},
  journal = {Optics Letters},
  year = 2008,
  volume = {33},
  number = {2},
  pages = {156--158},
  doi = {10.1364/OL.33.000156},
}

@INPROCEEDINGS{farneback2003two,
  author = {{Farneb{\"a}ck}, Gunnar},
  title = {Two-Frame Motion Estimation Based on Polynomial Expansion},
  booktitle = {Image Analysis, Scandinavian Conference on Image Analysis (SCIA)},
  series = {Lecture Notes in Computer Science},
  volume = {2749},
  pages = {363--370},
  year = 2003,
  publisher = {Springer},
  doi = {10.1007/3-540-45103-X_50},
}

@ARTICLE{theil1950rank,
  author = {{Theil}, Henri},
  title = {A rank-invariant method of linear and polynomial regression analysis},
  journal = {Proceedings of the Koninklijke Nederlandse Akademie van Wetenschappen},
  year = 1950,
  volume = {53},
  pages = {386--392},
}

@ARTICLE{neckel1984fts,
  author = {{Neckel}, Heinz and {Labs}, Dietrich},
  title = {The solar radiation between 3300 and 12500 {{\AA}}},
  journal = {Solar Physics},
  year = 1984,
  volume = {90},
  number = {2},
  pages = {205--258},
  doi = {10.1007/BF00173953},
}

@MISC{ispy2021,
  author = {{D{\'\i}az Baso}, Carlos and {Vissers}, Gregal and {Calvo}, Flavio and
            {Pietrow}, Alexander G. M. and {Yadav}, Rahul and {Robustini}, Carolina and
            {Rodr{\'\i}guez}, Jaime de la Cruz and {L{\"o}fdahl}, Mats},
  title = {{ISPy}: Solar Physics Tools in Python},
  year = 2021,
  howpublished = {Zenodo},
  doi = {10.5281/zenodo.5608440},
}

@ARTICLE{qiu2022calibration,
  author = {{Qiu}, Ye and {Rao}, Shihao and {Li}, Chuan and {Fang}, Cheng and
            {Ding}, Mingde and {Li}, Zhen and {You}, Wei and {Yuan}, Yuan and others},
  title = {Calibration procedures for the {CHASE/HIS} science data},
  journal = {Science China Physics, Mechanics, and Astronomy},
  year = 2022, volume = {65}, number = {8}, pages = {289603},
  doi = {10.1007/s11433-022-1900-5},
}

@INPROCEEDINGS{Baker1970,
  author = {{Baker}, Donald M.},
  title = {Flare classification based upon {X}-ray intensity},
  booktitle = {Observation and Prediction of Solar Activity Conference},
  year = 1970,
  publisher = {American Institute of Aeronautics and Astronautics},
  address = {Huntsville, AL},
  note = {AIAA Paper 70-1370},
  doi = {10.2514/6.1970-1370},
}

@ARTICLE{Li2023,
  author = {{Li}, Dong and {Li}, Chuan and {Qiu}, Ye and {Rao}, Shihao and
            {Warmuth}, Alexander and {Schuller}, Frederic and {Zhao}, Haisheng and
            {Shi}, Fanpeng and {Xu}, Jun and {Ning}, Zongjun},
  title = {Observational Signatures of Electron-driven Chromospheric Evaporation
           in a White-light Flare},
  journal = {The Astrophysical Journal},
  year = 2023,
  volume = {954},
  number = {1},
  pages = {7},
  doi = {10.3847/1538-4357/ace256},
}

@ARTICLE{Li2024,
  author = {{Li}, Qiao and {Li}, Ying and {Su}, Yang and {Song}, Dechao and
            {Li}, Hui and {Feng}, Li and {Huang}, Yu and {Li}, Youping and
            {Li}, Jingwei and {Zhao}, Jie and {Lu}, Lei and {Ying}, Beili and
            {Xue}, Jianchao and {Zhang}, Ping and {Tian}, Jun and {Liu}, Xiaofeng and
            {Li}, Gen and {Jing}, Zhichen and {Li}, Shuting and {Shi}, Guanglu and
            {Tian}, Zhengyuan and {Chen}, Wei and {Su}, Yingna and
            {Zhang}, Qingmin and {Li}, Dong and {Ge}, Yunyi and {Shan}, Jiahui and
            {Zhou}, Yue and {Lei}, Shijun and {Gan}, Weiqun},
  title = {Spectral and Imaging Observations of a {C2.3} White-Light Flare from the
           {Advanced} {Space}-{Based} {Solar} {Observatory} ({ASO-S}) and the {Chinese}
           {H}$\alpha$ {Solar} {Explorer} ({CHASE})},
  journal = {Solar Physics},
  year = 2024,
  volume = {299},
  number = {5},
  pages = {73},
  doi = {10.1007/s11207-024-02313-y},
}

@ARTICLE{Ma2024,
  author = {{Ma}, Y. L. and {Lao}, Q. H. and {Cheng}, X. and {Wang}, B. T. and
            {Zhao}, Z. H. and {Rao}, S. H. and {Li}, C. and {Ding}, M. D.},
  title = {{Sun}-as-a-star Study of an {X}-class Solar Flare with Spectroscopic
           Observations of {CHASE}},
  journal = {The Astrophysical Journal},
  year = 2024,
  volume = {966},
  number = {1},
  pages = {45},
  doi = {10.3847/1538-4357/ad3446},
}

@ARTICLE{Qiu2024,
  author = {{Qiu}, Ye and {Li}, Chuan and {Guo}, Yang and {Li}, Zhen and
            {Ding}, Mingde and {Kong}, Linggao},
  title = {Three-dimensional Velocity Fields of the Solar Filament Eruptions
           Detected by {CHASE}},
  journal = {The Astrophysical Journal Letters},
  year = 2024,
  volume = {961},
  number = {2},
  pages = {L30},
  doi = {10.3847/2041-8213/ad1e4f},
}

@ARTICLE{Liu2025,
  author = {{Liu}, Xiaofeng and {Hou}, Yijun and {Li}, Ying and {Qiu}, Ye and
            {Li}, Ting and {Cai}, Yingjie and {Rao}, Shihao and {Zhang}, Junyi and
            {Li}, Chuan},
  title = {{Sun}-as-a-star Analysis of the Solar Eruption Source Region Using
           {H}$\alpha$ Spectroscopic Observations of {CHASE}},
  journal = {The Astrophysical Journal},
  year = 2025,
  volume = {993},
  number = {1},
  pages = {126},
  doi = {10.3847/1538-4357/ae0743},
}

@ARTICLE{Rao2024,
  author = {{Rao}, Shihao and {Li}, Chuan and {Ding}, Mingde and {Hong}, Jie and
            {Chen}, Feng and {Fang}, Cheng and {Qiu}, Ye and {Li}, Zhen and
            {Chen}, Pengfei and {Li}, Kejun and {Hao}, Qi and {Guo}, Yang and
            {Cheng}, Xin and {Dai}, Yu and {Peng}, Zhixin and {You}, Wei and
            {Yuan}, Yuan},
  title = {Height-dependent differential rotation of the solar atmosphere
           detected by {CHASE}},
  journal = {Nature Astronomy},
  year = 2024,
  volume = {8},
  pages = {1102--1109},
  doi = {10.1038/s41550-024-02299-4},
}

@ARTICLE{Buil2023,
  author = {{Buil}, Christian and {Malherbe}, Jean-Marie and {Maksimovic}, Milan},
  title = {{Sol'Ex} et l'imagerie monochromatique solaire},
  journal = {Photoniques},
  year = 2023,
  number = {120},
  pages = {36--40},
  doi = {10.1051/photon/202312036},
}

@ARTICLE{Pal2025,
  author = {{V{\'a}radi Nagy}, P{\'a}l and {Pietrow}, A. G. M.},
  title = {An Atlas of Spectroheliograms from 3641 to 6600 {{\AA}}},
  journal = {Research Notes of the American Astronomical Society},
  year = 2025,
  volume = {9},
  number = {7},
  pages = {188},
  doi = {10.3847/2515-5172/adef50},
}

@MISC{Champeau2026,
  author = {{Champeau}, C{\'e}dric},
  title = {{JSol'Ex} ({melix/astro4j} 5.1.0): {Sol'Ex} data processing software},
  year = 2026,
  howpublished = {Zenodo},
  doi = {10.5281/zenodo.19582043},
}

@ARTICLE{Yu2024,
    author = {{Yu}, Fu and {Rao}, Shihao and {Zhao}, Jie and {Li}, Chuan and {Su}, Yang and {Zhao}, Jinsong and {Qiu}, Ye and {Ding}, Mingde and {Fang}, Cheng and {Li}, Hui and {Gan}, Weiqun},
    title = {Source Region of the Solar Wind: Statistics of the Doppler Velocities at the Chromosphere},
    journal = {The Astrophysical Journal Letters},
    year = 2024, volume = {968}, pages = {L20},
    doi = {10.3847/2041-8213/ad50c7},
  }

@ARTICLE{Wang2026,
    author = {{Wang}, Yue and {Li}, Chuan and {Rao}, Shihao and {Yu}, Haocheng and {Li}, Zhen and {Fang}, Cheng},
    title = {Revisiting {Evershed} Flows with {CHASE} Spectroscopic Observations},
    journal = {The Astrophysical Journal},
    year = 2026, volume = {999}, pages = {221},
    doi = {10.3847/1538-4357/ae3f9b},
  }

@ARTICLE{Xie2026,
    author = {{Xie}, T. Y. and {Zhao}, Z. H. and {Cheng}, X. and {Chen}, Y. H. and {Zheng}, Z. and {Hao}, Q. and {Li}, C. and {Ding}, M. D.},
    title = {Statistics of Solar Filament Mass Based on {CHASE} Sun-as-a-star Spectroscopic Observations},
    journal = {The Astrophysical Journal Letters},
    year = 2026, volume = {1008}, pages = {L45},
    doi = {10.3847/2041-8213/ae9850},
  }

@ARTICLE{Hong2022,
    author = {{Hong}, Jie and {Qiu}, Ye and {Hao}, Qi and {Xu}, Zhi and {Li}, Chuan and {Ding}, Mingde and {Fang}, Cheng},
    title = {Statistical analysis of the {Si} {I} 6560.58 \AA{} line observed by {CHASE}},
    journal = {Astronomy \& Astrophysics},
    year = 2022, volume = {668}, pages = {A9},
    doi = {10.1051/0004-6361/202244427},
  }

\end{document}